\documentclass{webofc}
\usepackage[varg]{txfonts}   
\usepackage{subfig}
\usepackage{siunitx}

\begin{document}
\title{Space-point calibration of the ALICE TPC with track residuals}
%
%

\author{\firstname{Marten Ole} \lastname{Schmidt}\inst{1}\fnsep\thanks{\email{oschmidt@physi.uni-heidelberg.de}} for the ALICE Collaboration
}

\institute{Physikalisches Institut, University of Heidelberg, Germany 
          }

\abstract{%
In the upcoming LHC Run 3, starting in 2021, the upgraded Time Projection Chamber (TPC) of the ALICE experiment will record minimum bias Pb--Pb collisions in a continuous readout mode at an interaction rate up to $\SI{50}{\kilo\hertz}$. This corresponds to typically 4-5 overlapping collisions during the electron drift time in the detector. Despite careful tuning of the new quadruple GEM-based readout chambers, which fulfill the design requirement of an ion backflow below 1\%, these conditions will lead to space-charge distortions of several centimeters that fluctuate in time. They will be corrected via a calibration procedure that uses the information of the Inner Tracking System (ITS), which is located inside, and the Transition Radiation Detector (TRD) and Time-Of-Flight system (TOF), located around the TPC, respectively. By using such a procedure the intrinsic track resolution of the TPC of a few hundred micrometers can be restored.

The required online tracking algorithm for the TRD, which is based on a Kalman filter, is presented. The procedure matches extrapolated ITS-TPC tracks to TRD space-points utilizing GPUs.
Subsequently these global tracks are refitted neglecting the TPC information. The residuals of the TPC clusters to the interpolation of the refitted tracks are used to create a map of space-charge distortions. Regular updates of the map compensate for changes in the TPC conditions. The map is applied in the final reconstruction of the data. First performance results of the tracking algorithm will be shown.

}
\maketitle
\section{Introduction}
\label{intro}
A Large Ion Collider Experiment (ALICE) is the dedicated heavy-ion experiment at the Large Hadron Collider (LHC) at CERN, designed to study the physics of strongly interacting matter at extreme energy densities and temperatures.
The ALICE apparatus consists of a central barrel enclosed by a solenoidal magnet and a muon spectrometer in the forward region. During the second LHC data taking period between 2015 and 2018 (commonly denoted as Run 2) the main tracking detector inside the central barrel, the TPC, was affected by large local distortions induced by space charge \cite{distortion-note}.
While the space-charge distortions were small in the bulk of the detector and $\mathcal{O}$(cm) only in localized regions they still needed to be corrected for in order to preserve the full detector performance.

In Run 3, ALICE will record Pb--Pb collisions in a continuous mode rather than a triggered, allowing for the readout of the full interaction rate up to $\SI{50}{\kilo\hertz}$, which is a factor six higher than the rate of Run 2. The MWPC-based readout chambers will be replaced by quadruple GEM-based chambers with an ion backflow below 1\% \cite{alicetpcupgrade}.
The distortions are expected to reach 10-$\SI{15}{\centi\meter}$ and be present in the whole TPC volume.

In addition to detector upgrades, an entirely new software framework (called $\mathrm{O}^2$) is being developed, where the functionalities of the data acquisition, High-Level Trigger (HLT), and the offline systems of the previous data collection runs will be combined \cite{aliceo2upgrade}. The space-charge distortion correction, which will be introduced in Sec.~\ref{calibration}, therefore needs to be ported to $\mathrm{O}^2$ and new requirements need to be incorporated. An important change is the revised readout of the TRD \cite{readout-tdr} which necessitates a new tracking procedure that is described in Sec.~\ref{algorithm}.

\section{Calibration procedure}
\label{calibration}

\begin{figure}
\centering
\sidecaption
\includegraphics[width=.6\textwidth]{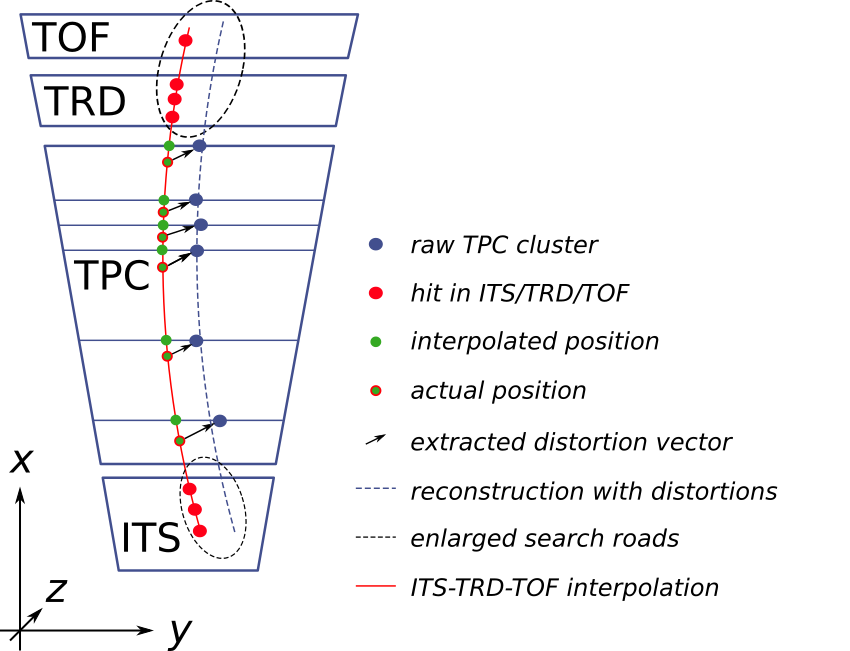}
\caption{Illustration of the space-point calibration for the TPC with track residuals. }
\label{fig-calibration}
\end{figure}
The space-point calibration of the TPC utilizes the external detectors ITS on the inside and TRD and TOF on the outside. Since they are not affected by space-charge distortions, they can provide reference cluster positions for global tracks inside the TPC. The calibration procedure is illustrated in Fig.~\ref{fig-calibration}. It consists of the following steps:
\begin{enumerate}
\item Track seeding and following in the TPC is done with relaxed tolerances.
\item Tracks are first matched to the ITS on one side, subsequently to the TRD and TOF on the other, again with relaxed tolerances.
\item Two independent refits are performed based on information from ITS and from TRD and TOF, respectively. In this step the TPC cluster information is ignored, but their association to the ITS-TRD-TOF tracks is kept.
\item The residuals between the distorted TPC clusters and the weighted ITS-TRD-TOF refits are collected for sub-volumes of the TPC.
\item For each sub-volume a vector representing the distortion in $x$, $y$ and $z$ is calculated.
\item A carefully tuned kernel smoother is applied to obtain a stable map of distortions without neglecting real discontinuities.
\end{enumerate}
One distortion map requires at least 450k matched tracks. In Run 2 one map was created about every 40 minutes. With the increased interaction rate and continuous readout in Run 3 around 1 minute of data taking will suffice.
The space-charge distortion maps correct the average TPC cluster positions. Fluctuations on a short time scale cannot be accounted for with the long integration time. They are covered by a second calibration procedure where the integrated charge on the readout pads will be sampled in intervals of about $\SI{1}{\milli\second}$ to allow for additional maps with a higher time granularity. The present work focuses on the correction of the average distortions.

The TPC seeding and track following required for step 1 is discussed in \cite{chep2018}. In the following the matching to the TRD which is needed for step 2 is described.

\section{TRD tracking algorithm}
\label{algorithm}
\begin{figure}
\centering
\sidecaption
\includegraphics[width=.6\textwidth]{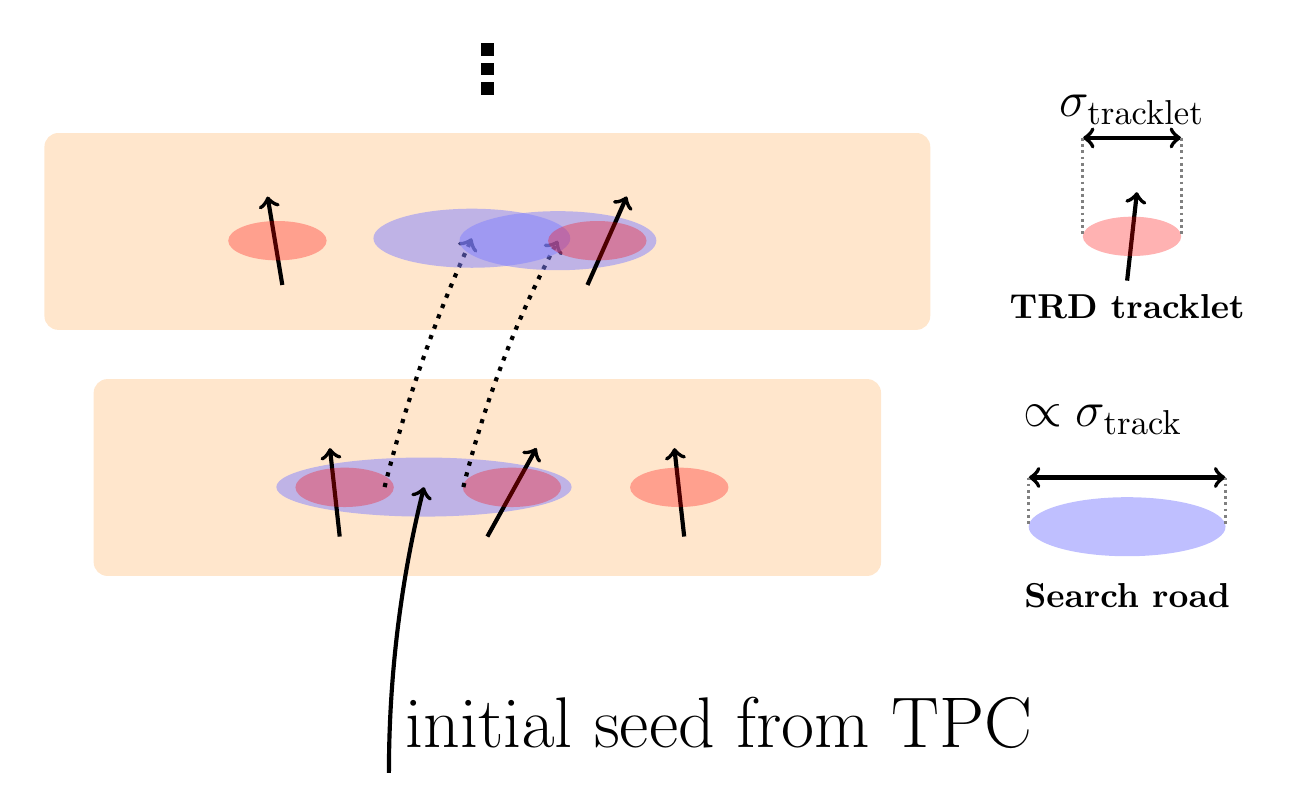}
\caption{Matching of TRD tracklets to extrapolated ITS-TPC tracks is using the Kalman filter technique. Multiple hypotheses can be kept per layer.}
\label{fig-algo}
\end{figure}

The TRD cannot be operated in continuous mode in Run 3 but will continue to be operated in triggered mode \cite{readout-tdr}. In order to utilize the available bandwidth in an optimal way, TRD readout will be limited to tracklets calculated in the detector Front-End Electronics (FEE). In Runs 1 and 2 these tracklets were used to generate triggers on high transverse momentum processes \cite{trd-paper}. The offline reconstruction was based on the charge cluster data which will not be available anymore in Run 3. Therefore, a new tracking algorithm based on tracklets is needed.

The TRD is segmented into 522 chambers. In azimuth the TRD follows the same segmentation as the TPC in 18 sectors. In the longitudinal direction 5 stacks are installed which each in turn consist of 6 layers in radial direction. A chamber consists of a radiator, followed by a $\SI{3}{\centi\meter}$ long drift region filled with Xe-$\mathrm{CO}_2$ and, in the same gas volume, a MWPC with pad readout. The signal on the readout pads is sampled in time bins of $\SI{100}{\nano\second}$. In the ideal case a traversing charged particle generates one cluster per time bin in a chamber.
The tracklets comprise results of a straight line fit $\tilde{y}(x) = y + (x - x_0) \cdot d_y$ to these clusters, where $y$ is the transverse offset with respect to the chamber center at the radius $x_0$ (close to the readout pad plane) and $d_y$ is the transverse deflection for a cluster over the full drift length. Additionally the longitudinal position $z$ (along the beams) in form of the pad row on which the tracklet was reconstructed and the PID information based on the energy loss in separate time windows in the detector are contained in a tracklet.

ITS-TPC matched tracks are extrapolated to the first TRD layer. All tracklets inside a search road based on the covariance matrices of the tracks are considered for matching. They are sorted depending on their $\chi^2$ with respect to the track. Only the offset of the straight line fit $y$ and the longitudinal position are used for the $\chi^2$ calculation, because the resolution of the tracklet inclination in azimuth ($\propto d_y$) is much worse compared to the precision for the inclination of the extrapolated tracks due to their short lever arm.
The tracklet inclination is instead used as criterion for exclusion if it deviates by more than $4\sigma$ from the track inclination. Multiple hypotheses can be kept per layer, but only the best $N_{\mathrm{candidates}}$ hypotheses are propagated as seeds to the next layer in order to limit combinatorics. The matching algorithm is depicted in Fig.~\ref{fig-algo}.

\begin{figure}
\centering
\subfloat{\includegraphics[width=.5\textwidth]{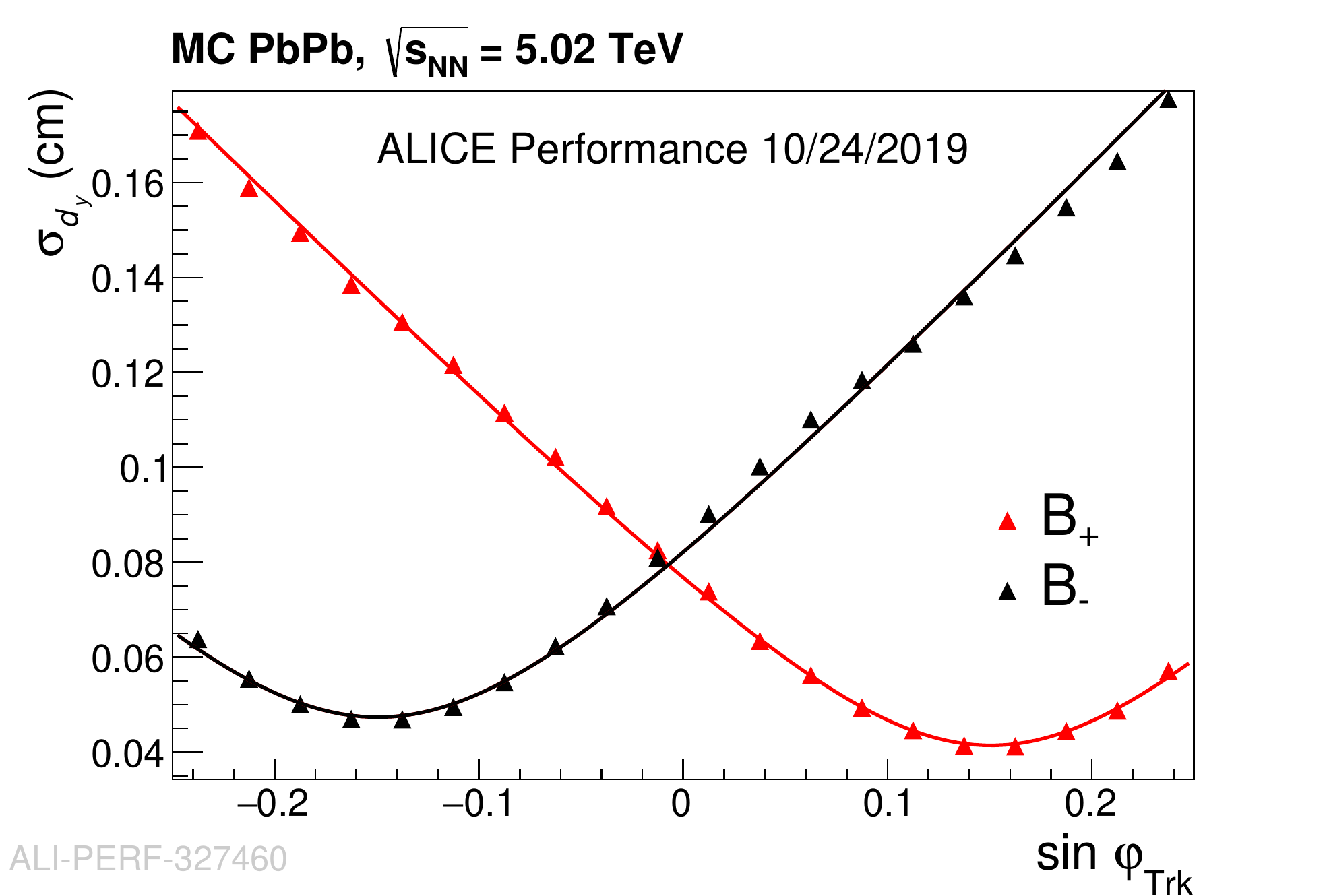}}
\subfloat{\includegraphics[width=.5\textwidth]{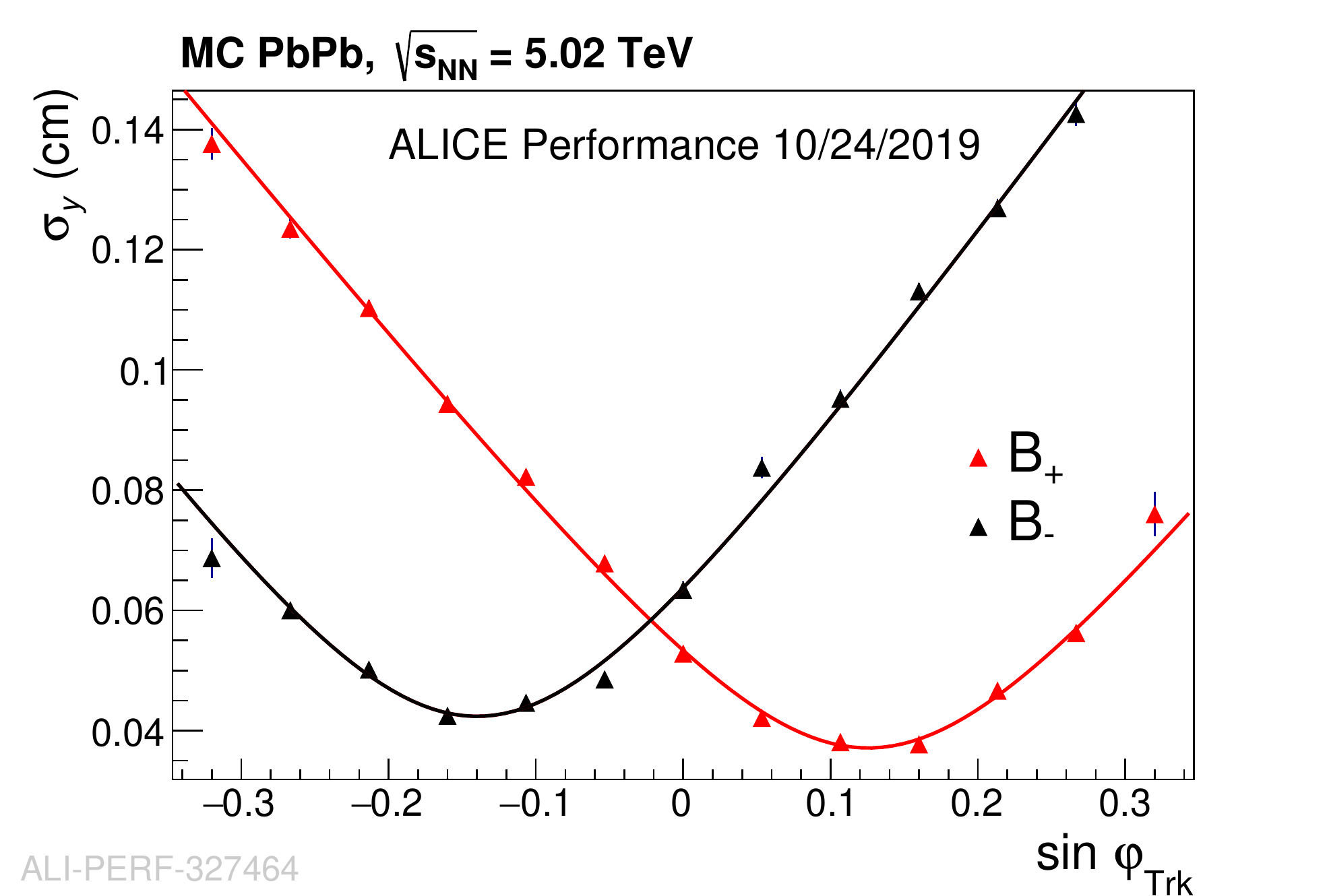}}
\caption{Resolutions for the straight line fit parameters contained in the tracklets. The accuracy of the deflection $d_y$ is shown on the left and the accuracy of the offset $y$ on the right. Optimal resolutions are given for tracks entering the detector under the Lorentz angle $\Psi_{\mathrm{Lorentz}} \approx \pm 8^{\circ}$. The sign depends on the polarity of the magnetic field $B_{\pm} = \pm \SI{0.5}{\tesla}$. For more details see text.}
\label{fig-resolution}
\end{figure}

The TRD tracklets do not contain information on the quality of the fit. The angular and position resolution have been parametrized in Fig.~\ref{fig-resolution}. A second order polynomial is fitted to the measured $d_y$ as a function of  the inclination of the ITS-TPC track in the azimuthal plane $\sin\varphi_{\mathrm{Trk}}$. The resolution $\sigma_{d_y}$ is defined as the width of the measured $d_y$ with respect to the fit for given track inclination. It can be directly translated into an angle with the fixed length of the drift region and is below $1^{\circ}$ for tracks with  $\varphi_{\mathrm{Trk}} \approx \Psi_{\mathrm{Lorentz}}$.
Since both resolutions depend on the inclination of the track associated to the tracklet, they have to be calculated on-the-fly during the matching procedure for the $\chi^2$ calculation, possibly multiple times if more than one track points in the same direction.

Due to misalignment not all tracklets are necessarily reconstructed at the same radius. This is handled by first extrapolating the track to the average radius of a chamber taking into account inhomogenities of the magnetic field and energy loss, and subsequently performing fast linear extrapolations for $\chi^2$ calculations.
Before the tracks are updated with the best matching tracklets they are again propagated with maximal precision to the exact radius of the matched tracklet.

Additional complications arise from the fact that the TRD readout pads are tilted by $\pm2^{\circ}$ in order to improve the resolution in longitudinal direction. The errors in azimuth and longitude are therefore correlated and $y$ has to be corrected for based on the $z$ position of the extrapolated track.

\subsection{Performance and benchmark results}
\label{performance}
\begin{figure}
\centering
\subfloat{\includegraphics[width=.5\textwidth]{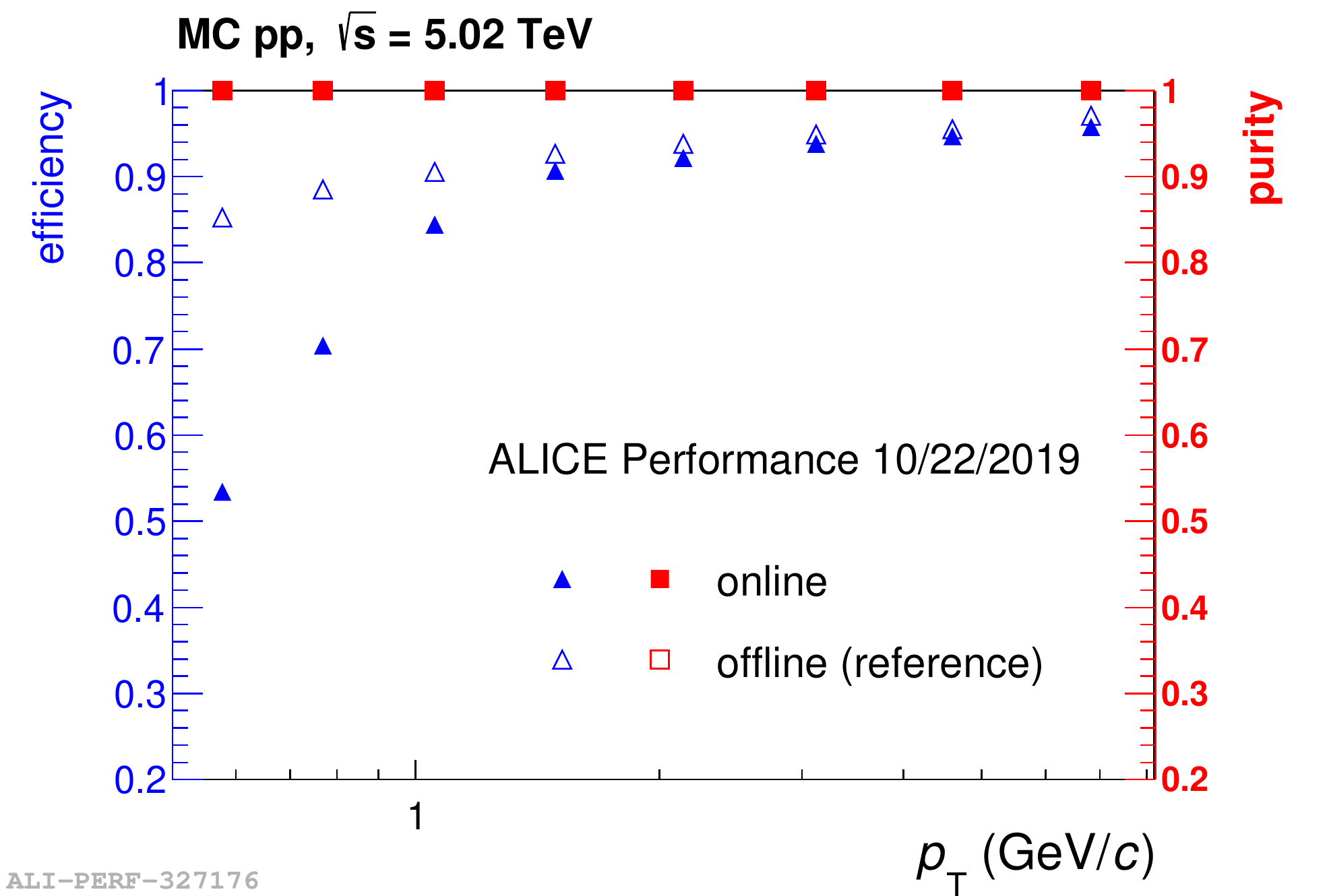}}
\subfloat{\includegraphics[width=.5\textwidth]{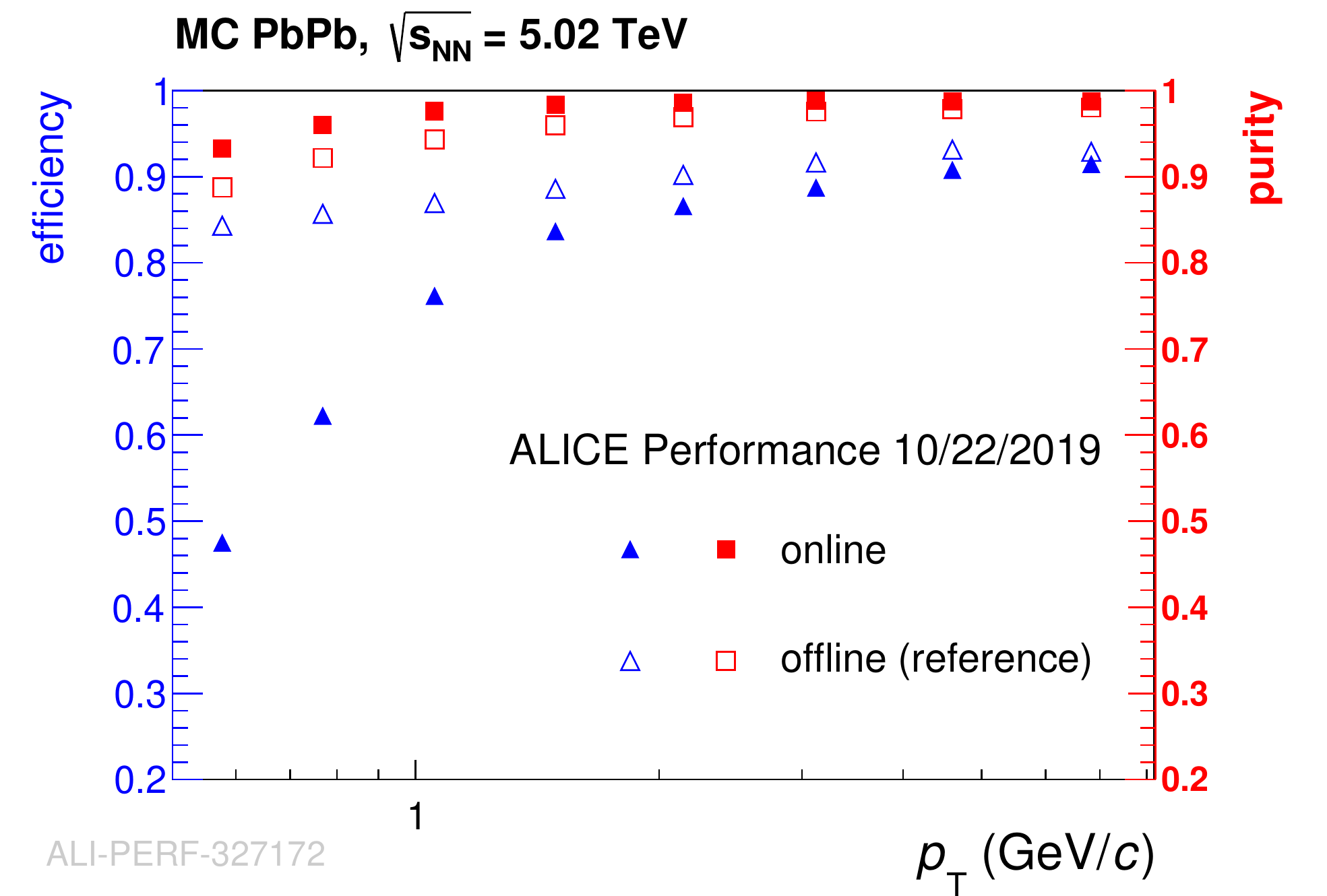}}
\caption{Comparison of the new online tracklet-based TRD tracking algorithm (closed symbols) with the cluster-based offline tracking algorithm (open symbols) as reference. Efficiency and purity are plotted both for pp collisions (left) and Pb--Pb collisions (right).}
\label{fig-performance}
\end{figure}
Efficiency and purity distributions for the TRD tracking are shown in Fig.~\ref{fig-performance}. Since a minimum of two tracklets are required for the TPC space-point calibration, the efficiency is defined as the fraction of tracks which are matched to at least 2 correct (based on their MC label) tracklets. Below $p_{\mathrm{T}} \approx \SI{1.5}{\giga\electronvolt}/c$ the efficiency for the new TRD tracking algorithm deteriorates compared to the old tracking, which is based on cluster data. This is due to the fact that the tracklets in the FEE are allowed to span over a maximum of two pads, introducing a selection on the deflection inside the drift region. This in turn translates into a position dependent selection on the transverse momentum which is reflected as the observed smeared decrease of efficiency. Primary tracks with $p_{\mathrm{T}} < \SI{300}{\mega\electronvolt}/c$ do not reach the TRD and are neglected by the tracking algorithm. At high-$p_{\mathrm{T}}$ the new tracking algorithm yields results compatible to the old offline tracking.

The purity is defined as the fraction of tracks with at least two attached tracklets which in addition do not have any fake tracklets attached. In pp collisions the purity for both the old and the new tracking algorithm is practically 1, while it decreases to about 0.9 for low momentum tracks in Pb--Pb collisions where the track density is much higher.

For the TPC space-point calibration a lower efficiency can be compensated by accumulating more statistics. The purity on the other hand is improved by neglecting central Pb--Pb events, where the track density is very high, when creating the distortion maps.

\begin{figure}
\centering
\includegraphics[width=\textwidth]{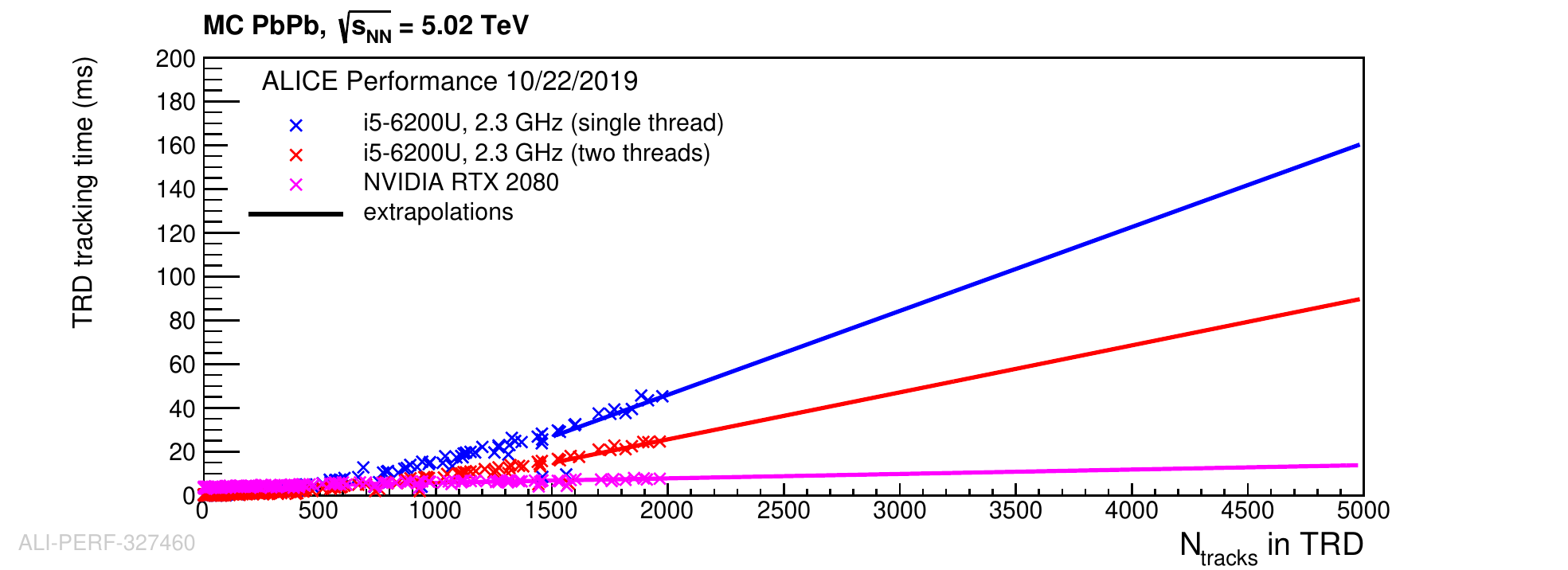}
\caption{Benchmark results for single-threaded, multi-threaded and GPU reconstructions.}
\label{fig-benchmark}
\end{figure}

Since the TRD tracking will run synchronously to the data taking the processing time is very important. Parallelization can easily be achieved over the tracks and is implemented both on CPUs via OpenMP and on NVIDIA GPUs via CUDA. The processing time per event as a function of the number of tracks which reach the TRD is shown in Fig.~\ref{fig-benchmark}. The most central events typically contain about 2000 ITS-TPC tracks which reach the TRD. This number is not yet large enough to profit strongly from GPU utilization.
For Run 3 the data will not be processed on a per-event basis, but in time frames of 10-$\SI{20}{\milli\second}$ corresponding to about 500-1000 collisions. The number of tracks to be processed is therefore much higher and large speedup on GPUs compared to CPUs is expected as can be deduced from the extrapolations shown in Fig.~\ref{fig-benchmark}.

\section{Next steps}
\label{steps}

Currently the simulation chain of the TRD in the new $\mathrm{O}^2$ framework for Run 3 is nearly complete. As soon as it is finished, the TRD tracking can be plugged into $\mathrm{O}^2$.
Furthermore the tracking algorithm currently supports only a reconstruction on a per-event basis. Support for tracking in time frames still needs to be added. Note that this does not increase the complexity, as the TRD will continue to be operated in triggered mode and therefore the association of tracklets to a certain bunch crossing is a priori known. After matching to ITS, the TPC tracks also contain a time stamp which connects them to a bunch crossing. Hence, only the sizes of the input arrays for tracklets and tracks increase. And for the matching only those tracklets which belong to the same interaction as the respective ITS-TPC track will be considered.

\section{Conclusions}
\label{conclusions}

The new matching between ITS-TPC tracks and TRD tracklets shows a comparable performance with the offline tracking of Run 1 and 2, which was based on the TRD raw data. At the same time it fulfills the computing speed requirements for Run 3. It was tested successfully both on CPUs and NVIDIA GPUs. The CPU version was activated in the HLT during data taking in 2018 both in pp and in Pb--Pb collisions.
Built on the ALICE GPU framework \cite{ieee2017} the algorithm is contained in a single source code file and supports wrappers to different APIs, i.e. there is no code duplication needed for the CPU and the GPU version.
The remaining steps specified in Sec.~\ref{calibration}, which are required to create the space-charge distortion maps for the TPC, have already been ported to the new $\mathrm{O}^2$ framework.
As soon as the TRD simulation chain is ready the full calibration can run in the Run 3 software framework.

%
%
%

\end{document}